\documentclass[]{ICCFD6}

\usepackage{latexsym}
\usepackage{epsfig}
\usepackage{fancyhdr}
\usepackage{float}
\usepackage{amsmath}

\usepackage[C40,T1]{CJKutf8}
\newcommand{\twn}[1]{\bgroup\begin{CJK*}{UTF8}{}\CJKtilde\CJKfamily{bkai}{#1}\end{CJK*}\egroup}

\newcommand{\LJeps}{\ensuremath{\epsilon}}
\newcommand{\LJsig}{\ensuremath{\sigma}}
\newcommand{\mardyn}{\textit{ls1 mardyn}}

\newcommand{\radius}{\ensuremath{r}}

\newcommand{\cutoff}{\ensuremath{\distance_\mathrm{c}}}
\newcommand{\degree}[1]{\ensuremath{{#1}^{\circ}}}

\newcommand{\distance}{\radius}
\newcommand{\FWenergy}{\ensuremath{\mathnormal{W}}}

\newcommand{\LJenergy}{\LJeps}
\newcommand{\LJenergyFW}{\ensuremath{\LJenergy_\mathrm{fw}}}
\newcommand{\LJsize}{\LJsig}
\newcommand{\LJsizeFW}{\ensuremath{\LJsize_\mathrm{fw}}}
\newcommand{\LJTS}{LJTS}

\newcommand{\vapour}{vapor}

\begin{document}

\title{Research on the behavior of liquid\\ fluids atop superhydrophobic  
gas-bubbled surfaces}

\author{Gerrit C.\ Lehmann, Frithjof Dubberke, Martin Horsch, Yow-Lin Huang (\twn{黃佑霖}),\\ Svetlana Miroshnichenko, R\"udiger Pflock, Gerrit Sonnenrein \& Jadran Vrabec\footnote{Corresponding author: Prof.\ Dr.-Ing.\ habil.\ Jadran Vrabec $<$jadran.vrabec@upb.de$>$, phone +49 5251 602 421.}}

\addr{Thermodynamics and Energy Technology Laboratory (ThEt),\\
University of Paderborn, Warburger Str.\ 100, 33098 Paderborn, Germany}


\abstract{{\bf Abstract}: Superhydrophobic surfaces play an important  
role in the development of new product coatings such as cars, but also  
in mechanical engineering, especially design of turbines and  
compressors. Thus a vital part of the design of these surfaces is the  
computational simulation of such with a special interest on variation  
of shape and size of minor pits grooved into plane surfaces.
In the present work, the dependence of the contact angle on the
fluid-wall dispersive energy is determined by molecular simulation
and static as well as dynamic properties of unpolar fluids in contact
with extremely rough surfaces are obtained.}

\maketitle

\pagestyle{empty}

%

\thispagestyle{empty}

\noindent
Fluid flow over extremely rough surfaces is governed by non-trivial
boundary conditions which can be related to the contact angle as discussed
by Voronov et al.\ (2008). Boundary slip is most relevant
for microscopic and nanoscopic flow, while the influence of
surface roughness on the contact angle becomes extreme in case of superhydrophobic surfaces.
For nanoscopic channel dimensions as well as roughness on the molecular length
scale, the accuracy of simulation results can be optimized by applying
molecular dynamics (MD), since this approach reflects the actual structure
of the material more directly than higher-level methods that rely on aggregated
models and properties.

As long as no
hydrogen bonds are formed between the wall and the fluid,
the interfacial properties mainly depend on the fluid-wall dispersive interaction,
even for hydrogen bonding fluids.
The truncated and shifted Lennard-Jones (\LJTS) potential
with a cutoff radius of $\cutoff$ = 2.5 $\LJsize$
accurately reproduces the dispersive interaction
if adequate values for the size and energy parameters $\LJsize$ and $\LJenergy$
are speci\-fied, cf.\ Vrabec et al.\ (2006).

\begin{figure}[t!]
\centering
\includegraphics[width=7.25cm]{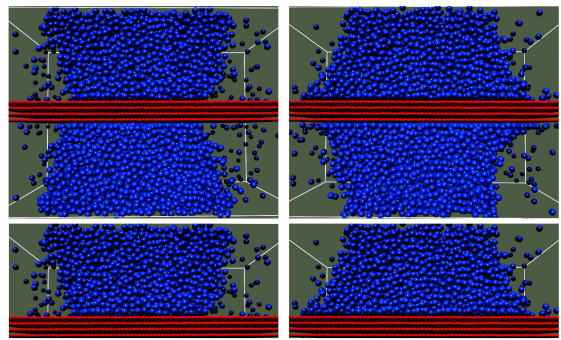}
\caption{
   Simulation snaphots for a smooth surface with a reduced fluid-wall
   dispersive energy $\FWenergy$ of 0.09 (left) and 0.16 (right) at a temperature
   of 0.73 $\epsilon / k$. The upper half is reproduced in the bottom to
   illustrate the effect of the periodic boundary condition.
}
\label{poehlmann}
\end{figure}

Fluid-wall interactions can be represented by Lennard-Jones-12-6 
effective potentials, acting between fluid particles and the atoms of
the solid, cf.\ Battezzati et al.\ (1975).
Following this approach, the \LJTS{} potential with the size and
energy parameters $\LJsizeFW = \LJsize$ as well as
$\LJenergyFW = \FWenergy\LJenergy$
was applied for the unlike interaction
using the same cutoff radius as for the fluid.
The wall was modeled as a system of coupled harmonic oscillators with different
spring constants for transverse and longitudinal motion,
adjusted to simulation results for graphite with a rescaled variant of
the Tersoff (1988) potential.
Massively parallel MD simulations were conducted with the
program \mardyn, cf.\ Bernreuther et al.\ (2009).
A periodic boundary condition was applied to the system, leaving a channel
with a diameter of 27 $\LJsize$ between the wall and its periodic image, cf.\
Fig.\ \ref{poehlmann}.

The contact angle was determined from the density profiles by averaging
over at least 800 ps after equilibration.
A circle was adjusted to the positions of the interface in the bins
corresponding to distances between 3 and 11 $\LJsize$ from the wall,
and the tangent to this circle at a distance of 1 $\LJsize$ from
the wall was consistently used to determine the contact angle.

\begin{figure}[h!]
\centering
\includegraphics[width=7.25cm]{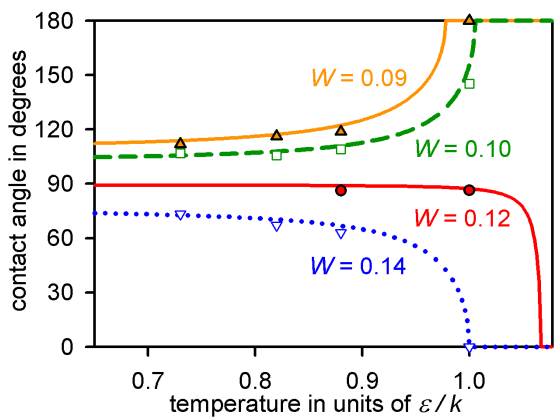}
\caption{
MD simulation results and correlation for the contact angle of the LJTS fluid on
a smooth surface in dependence of the temperature with
reduced fluid-wall dispersive energy $\FWenergy$ values
of 0.09 ($\Delta$ / ---), 0.10 (\square / -- --),
0.12 ($\bullet$ / ---) as well as 0.14 ($\nabla$ / $\cdot \cdot \cdot$).
The entire range between triple point and critical temperature is shown.
}
\label{fig11}
\end{figure}

A contact angle -- as opposed to total dewetting or wetting --
appears only for a relatively narrow range of $\FWenergy$ values.
As the temperature increases and the \vapour-liquid
surface tension decreases, the contact angle reaches more extreme values,
leading to the well-known phenomenon characterized
by Cahn (1977) as crticial point wetting, cf.\ Fig.\ \ref{fig11}.
This plot agrees quali\-tatively
with the results of Giovambattista et al.\ (2007) regarding the influence
of the polarity of hydroxylated silica surfaces on the contact angle formed
with water.

\begin{figure}[b!]
\centering
\includegraphics[width=7.25cm]{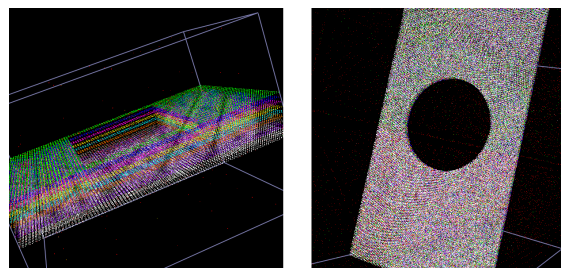}
\caption{
   Left: Rectangular elementary cell of a pit grid with  
   rectangular pit for simulation with gaseous and liquid fluids.
   Right: Rectangular elementary cell (prototype  
   version) with cylindrical bore for simulation of streaming fluids.
}
\label{gerrit_lehmann}
\end{figure}

For a constant value $\FWenergy = 0.09$ of the reduced fluid-wall energy,
corresponding to a contact angle of about $\degree{110}$ for moderate as
well as low temperatures, the surface shape and roughness
was varied in further simulations, cf.\ Fig.\ \ref{gerrit_lehmann}.
The stability of the Cassie state as well as the influence of
the surface shape on dynamic properties such as the boundary slip length
and slip velocity in nanoscopic Poiseuille flow were
studied by MD simulation. 

The simulation results regard the length scale
between 1 and 100 nm and can be reliably extrapolated to the characteristic
system dimensions corresponding to typical superhydrophobic surfaces, e.g.\
about one micron in case of the material manufactured 
by Steinberger et al.\ (2008). Thereby, the experimental point of
view can be complemented by a theoretical treatment, applying the 
variant of computational fluid dynamics that is best suited for
the investigation of nanopatterned surfaces: MD simulation.

The authors would like to thank the German Science Foundation (DFG)
for funding SFB 716 and M.\ Heitzig (Copen\-hagen), J.\ Harting
(Eind\-hoven), and D.\ Vollmer (Mainz) as well as M.\ Bernreuther, C.\ Dan,
and M.\ Hecht (Stuttgart) for technical support and fruitful
discussions. The presented research was carried out under the
auspices of the Boltzmann-Zuse Society of Computational Molecular
Engineering (BZS) and the simulations were performed on the
\textit{XC 4000} supercomputer at the Steinbuch Centre of Computing, Karlsruhe,
under the grant LAMO.

\end{document}